\documentclass{ws-procs9x6}
\begin{document}
\title{QUARK HELICITY FLIP AND THE TRANSVERSE SPIN DEPENDENCE 
OF INCLUSIVE DIS\footnote{Supported in part by DOE, and by NSF under 
grants PHY--0114343 and PHY--0301841. Notice: Authored by Jefferson Science 
Associates, LLC under U.S.\ DOE Contract No.~DE-AC05-06OR23177. 
The U.S.\ Government retains a non-exclusive, paid-up, irrevocable, 
world-wide license to publish or reproduce this manuscript for 
U.S.\ Government purposes.}}
\author{A.~AFANASEV$^{1,2}$, M.~STRIKMAN$^{3}$, C.~WEISS$^{2}$} 
\address{$^{1}$Department of Physics, Hampton University, Hampton, 
VA 23668, USA \\
$^{2}$ Theory Center, Jefferson Lab, Newport News, VA 23606, USA \\
$^{3}$ Dept.\ of Physics, Pennsylvania State University, 
University Park, PA 16802, USA}
\begin{abstract}
Inclusive DIS with unpolarized beam exhibits a subtle dependence on 
the transverse target spin, arising from the interference of one--photon 
and two--photon exchange amplitudes in the cross section. We argue that 
this observable probes mainly the quark helicity--flip amplitudes 
induced by the non-perturbative vacuum structure of QCD (spontaneous
chiral symmetry breaking). This conjecture is based on (a)~the absence
of significant Sudakov suppression of the helicity--flip process
if soft gluon emission in the quark subprocess is limited by
the chiral symmetry--breaking scale $\mu^2_{\text{chiral}} 
\gg \Lambda^2_{\textrm{QCD}}$; (b)~the expectation that the quark
helicity--conserving twist--3 contribution is small.
The normal target spin asymmetry is estimated to be of the order 
$10^{-4}$ in the kinematics of the planned Jefferson Lab Hall~A experiment. 
\\
\end{abstract}
%
%
%
%
\bodymatter
A fundamental property of QCD is that the quark helicity is conserved
in perturbative interactions, because of chiral invariance in the
light--quark sector. It is known, however, that non-perturbative
interactions at distances of the order $1/\mu_{\textrm{chiral}}
\sim 0.3 \, \text{fm}$ cause the dynamical breaking
of chiral symmetry and lead to the appearance of non-zero quark 
helicity--flip amplitudes. This effect essentially determines the
structure and interaction of hadrons at large distances. An interesting 
question is what the presence of quark helicity--flip amplitudes
implies for hard electromagnetic processes (invariant momentum 
transfer $Q^2 \gg 1 \, \textrm{GeV}^2$). Such processes couple directly to the 
quark degrees of freedom and in certain cases can be described by 
factorization of the cross section/amplitude into ``hard'' and ``soft'' 
contributions. Quark helicity flip has been discussed
{\it e.g.}\ in relation to the high--$Q^2$ behavior of the proton 
form factor ratio $Q F_2 / F_1$ and wide--angle Compton 
scattering \cite{Miller:2002qb,Miller:2004rc}.
We have recently pointed out\cite{Afanasev:2007ii} 
that quark helicity flip in hard processes
can be probed in the transverse target spin dependence of inclusive 
deep--inelastic scattering (DIS) with unpolarized beam, 
$eN \rightarrow e'X$. In these proceedings 
we summarize the arguments relating this observable to quark 
helicity flip and discuss our estimate of the magnitude of the 
expected asymmetry (for details, see Ref.~\refcite{Afanasev:2007ii}).

%
%
\begin{figure}[t]
\includegraphics[width=0.98\textwidth]{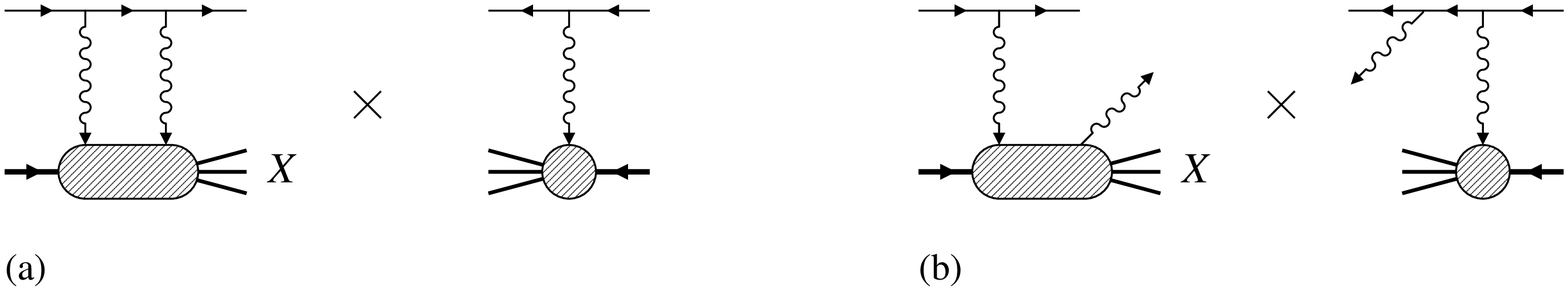}
\caption[]{QED processes contributing to the transverse
target spin dependence of the inclusive $eN$ cross section at $O(\alpha^3 )$.
(a)~Interference of one--photon and two--photon exchange.
(b)~Real photon emission (bremsstrahlung).}
\label{fig:twogamma}
\end{figure}
The differential cross section of inclusive $eN$ scattering with 
unpolarized beam depends on the target spin $\bm{S}$ as
\begin{equation}
d\sigma \;\; = \;\;
d\sigma_U  \; + \;  
\frac{(\bm{S}, \bm{l} \times \bm{l}')}{|\bm{l} \times \bm{l}'|}
\; d\sigma_N 
\label{dsigma}
\end{equation}
($\bm{l}, \bm{l}'$ are the initial/final electron momenta in the lab frame),
and the normal spin asymmetry is defined as
\begin{equation}
A_N \;\; \equiv \;\; \frac{d\sigma_N}{d\sigma_U} .
\label{A_N}
\end{equation}
The spin--dependent term is zero in the one--photon exchange approximation 
to DIS, by the combination of $P$ and $T$ invariance and the hermiticity 
of the electromagnetic current operator (Christ--Lee theorem).
A non-zero spin dependence appears only in higher orders of the QED
coupling constant, from the interference of one--photon and absorptive 
two--photon exchange amplitudes in the cross section 
(Fig.~\ref{fig:twogamma}a), as well as
from real photon emission (bremsstrahlung, Fig.~\ref{fig:twogamma}b).
It is important that the two contributions to the spin--dependent 
cross section are individually free of QED infrared (IR) divergences 
and thus can be considered as distinct physical effects; this follows 
from the general factorization properties and the spin--independence 
of IR photon exchange in QED \cite{Afanasev:2007ii}. 
The two--photon exchange amplitude is also free of QED collinear 
divergences (vanishing photon virtuality at non-zero momentum); 
this is a general consequence of electromagnetic
gauge invariance \cite{Afanasev:2007ii,Afanasev:2004hp}. 
The spin--dependent two--photon 
exchange cross section, Fig.~\ref{fig:twogamma}a, thus represents a
well--defined observable in QED, which can be used as a new probe
of hadron structure.

%
%
\begin{figure}[t]
\includegraphics[width=0.9\textwidth]{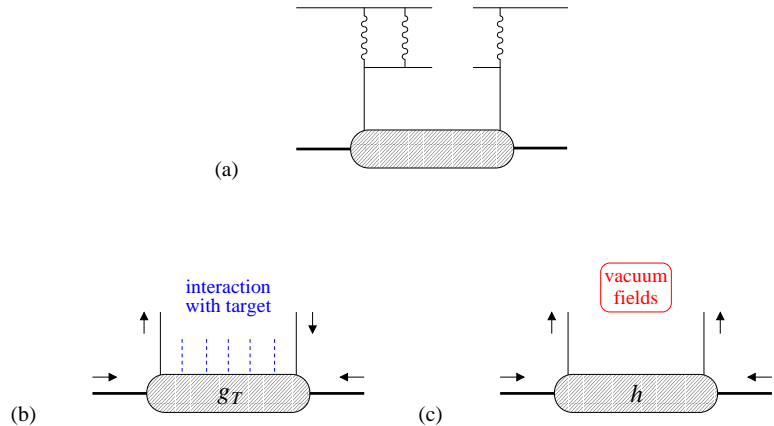}
\vspace{2ex}
\caption[]{Transverse spin dependence of the DIS cross section 
in QCD. (a) Dominance of two--photon exchange 
with the same quark. (b) Quark helicity--conserving process
involving interactions with the target remnants. 
(c) Quark helicity flip due to interaction with non--perturbative
vacuum fields.}
\label{fig:flip}
\end{figure}
In DIS kinematics we expect the spin--dependent two--photon exchange 
cross section to arise predominantly from the amplitudes in which
the two--photon exchange couples to a single quark, namely the
same quark which is hit in the interfering one--photon exchange 
amplitude, see Fig.~\ref{fig:flip}a. This expectation is based on
the absence of ``anomalous'' (IR or collinear--enhanced) contributions 
in the two--photon exchange amplitude, and on general considerations
of scattering from a hadronic system with momentum transfer 
$Q^2 \gg R_{\text{hadron}}^{-2}$. However, one can easily see that
no transverse spin dependence can arise in the parton model approximation
in which the electron scatters from an on--shell massless quark, as the 
transverse spin--dependent cross section for scattering from an on--shell 
quark requires quark helicity flip and is explicitly proportional to 
the quark mass. A spin dependence of the hadronic cross section 
thus can come only from the two mechanisms indicated
in Figs.~\ref{fig:flip}b and~c.

In the process of Fig.~\ref{fig:flip}b the quark helicity is conserved
in the electron--quark subprocess, and the helicity flip between the
interfering hadronic amplitudes happens at the level of the quark
distribution in the nucleon. This process requires non-zero virtuality
(off--shellness) of the active quark, and, at the same time, explicit
interaction of the active quark with the spectator system. The two
effects are linked by electromagnetic gauge invariance; only the
sum of the two maintains transversality of the amplitude and avoids
the appearance of unphysical collinear divergences. (An attempt to 
calculate the transverse spin asymmetry including finite virtuality
of the active quark but neglecting explicit interactions has produced a 
divergent result \cite{Metz:2006pe}.) In the process of Fig.~\ref{fig:flip}c 
the quark helicity is flipped in the electron--quark subprocess.
This process requires interaction of the active quark with the 
non-perturbative vacuum fields which cause the spontaneous breaking of
chiral symmetry. It does not require interactions with the spectator
system and exists already for on--shell quarks. Perturbative QCD 
interactions do not ``mix'' the contributions of Figs.~\ref{fig:flip}b and c.
An interesting question is whether in typical DIS kinematics 
one of the two mechanisms dominates or both make comparable
contributions to the transverse spin dependence.

Following Ref.~\refcite{Afanasev:2007ii}, we suggest here that the quark 
helicity--flip contribution, Fig.~\ref{fig:flip}c, may be sizable and 
could well be the dominant mechanism in the transverse target spin
dependence. This perhaps somewhat surprising assertion rests on 
the following two observations.

(1) \textit{Insignificant Sudakov suppression of 
helicity--flip process.} The helicity--flip process in QCD requires 
propagation through quark configurations with virtualities 
$\lesssim \mu^2_{\textrm{chiral}}$, which experience significant 
interactions with the chiral symmetry breaking vacuum fields.
This leads to Sudakov suppression of the cross section compared 
to the case of unrestricted virtualities. A similar suppression takes 
place in the usual DIS cross section (where the restriction to 
small virtualities results from the condition of producing real 
particles in the final state); however, there it is compensated by 
real gluon emissions. In the case of the transverse spin asymmetry
this compensation is likely to be incomplete, and some residual
Sudakov suppression of the interference cross section should remain.
Numerical estimates based on the on--shell Sudakov formfactor
of QCD show that this suppression should be marginal for 
$Q^2 \sim \text{few GeV}^2$, if the IR cutoff governing soft gluon
emission in the Sudakov formfactor is of the order of 
$\mu^2_{\textrm{chiral}} \gg \Lambda_{\textrm{QCD}}^2$.
While we cannot presently prove that this magnitude of the IR cutoff
is dictated by chiral symmetry breaking in QCD, it certainly appears 
natural in the light of dynamical models such as the instanton vacuum, 
which suppose that gluons of wavelength $< \mu_{\textrm{chiral}}^{-1}$ 
are ``contained'' in the non-perturbative field configurations 
which break chiral symmetry.

\begin{figure}[t]
\begin{center}
\includegraphics[width=0.5\textwidth]{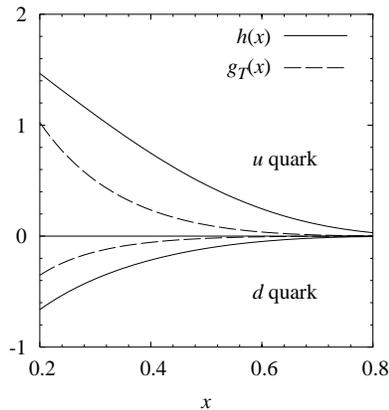}
\end{center}
\caption[]{Comparison of the proton's twist--3 helicity--conserving 
quark distribution $g_T (x)$ [calculated in terms of the twist--2
quark helicity distribution $g(x)$ using the Wandzura--Wilczek 
relation] with the twist--2 helicity--flip (transversity) distribution
$h(x)$ [estimated assuming that $h(x) = g(x)$, \textit{i.e.}, 
transversity = helicity distribution]. 
Shown are the results for both $u$ and $d$ quarks.}
\label{fig:ww}
\end{figure}

(2) \textit{Non-partonic character of helicity--conserving
process.} The quark helicity--conserving process, Fig.~\ref{fig:flip}b,
explicitly involves non-zero virtuality of the initial quark and its
interaction with the spectator system, and is thus of essentially
``non-partonic'' character. The calculation of this process within
the collinear QCD expansion starts from the ``handbag diagram'' with
the twist--3 quark helicity--conserving distribution $g_T(x)$, 
which also appears in the calculation of the spin structure function 
$g_1 + g_2$ measured with polarized beam and transversely polarized target. 
However, retaining only the contribution from the ``handbag diagram'' 
would not be a consistent approximation for the spin--dependent 
two--photon exchange cross section, as this would violate electromagnetic
gauge invariance and lead to the appearance of QED collinear divergences.
The interaction of the active quark with the gluon field in the target
needs to be included in order to restore electromagnetic gauge invariance.
An interesting question is whether this ``restoration'' of gauge invariance 
will altogether eliminate the contribution from the dynamical twist--2 
operators originally present in $g_T$, or whether some part of it
survives in the final result. In the former case the helicity--conserving
contribution to the asymmetry would be governed by the same mechanism
as the twist--3 (``non--Wandzura--Wilczek'') part of $g_2$, 
which has been shown to be very small compared to $g_1$ 
by the SLAC \cite{Anthony:2002hy} and Jefferson Lab Hall~A 
\cite{Zheng:2004ce} experiments. In the latter case, one
could estimate the relative order--of--magnitude of the 
contributions from Figs.~\ref{fig:flip}b and c by comparing
\cite{Afanasev:2007ii}
\begin{equation}
\frac{\langle \bm{k}_T^2 \rangle}{M} \; g_{T, f}(x) 
\;\; \longleftrightarrow \;\; 
M_q \, h_f(x) ,
\end{equation}
where on the left--hand side 
$\langle \bm{k}_T^2 \rangle$ is the typical 
quark intrinsic transverse momentum, $M$ the nucleon mass, 
and $g_{T, f}(x)$ is given in terms of the twist--2 quark 
helicity distribution, $g_f(x)$, by the Wandzura--Wilczek relation,
and on the right--hand side $M_q \approx 0.3 \textrm{--} 0.4 \, 
\textrm{GeV}$ is a typical constituent quark mass, determining the strength 
of the helicity--flip amplitude for low--virtuality quarks, and $h_f$ the 
quark transversity distribution ($f = u, d$ denotes the quark flavor). 
The numerical estimate of Fig.~\ref{fig:ww} shows that even under 
these assumptions the quark helicity--conserving process is unlikely 
to dominate over the quark helicity--flip one.
%
%

To summarize, we have argued that a sizable, perhaps dominant,
contribution to the transverse spin--dependent cross section
comes from the quark helicity--flip process governed by the 
quark transversity distribution in the nucleon. To validate this 
conjecture, a more detailed investigation of the effects of
electromagnetic gauge invariance on the quark helicity--conserving
contribution is needed. We note that this problem bears some
similarity to that of gauge invariance in the QCD light--cone
expansion of deeply--virtual Compton scattering (DVCS), 
where it was found that the twist--2 contribution
to the amplitude alone is not transverse, and that transversality
is restored by including certain ``kinematical'' twist--3 
contributions \cite{Radyushkin:2000jy,Radyushkin:2000ap,Penttinen:2000dg}.
If the smallness of the helicity--conserving contribution 
of Figs.~\ref{fig:flip}b could indeed be confirmed by explicit calculation, 
the normal spin asymmetry in inclusive DIS, Eq.~(\ref{A_N}), would be 
a very interesting observable for testing the mechanism of 
non-perturbative quark helicity flip in QCD.

A numerical estimate of the normal spin asymmetry in DIS kinematics
was made in Ref.~\refcite{Afanasev:2007ii} using a light--front constituent
quark model, in which DIS is described as elastic scattering from 
pointlike constituent quarks, and quark helicity--flip amplitudes
are generated by the constituent quark mass. In order to have
a self--consistent picture, this model was treated in the
``composite nucleon'' approximation, where we suppose that the
quark transverse momenta, which are of the order of the inverse
nucleon size, are parametrically small compared to the constituent
quark mass,
\begin{equation}
\langle \bm{k}_T^2 \rangle \;\; \sim \;\; R_N^{-2} \;\; \ll \;\; M_q^2 . 
\label{neglect_k_T}
\end{equation}
%
%
\begin{figure}[t]
\begin{tabular}{cc}
\includegraphics[width=0.49\textwidth]{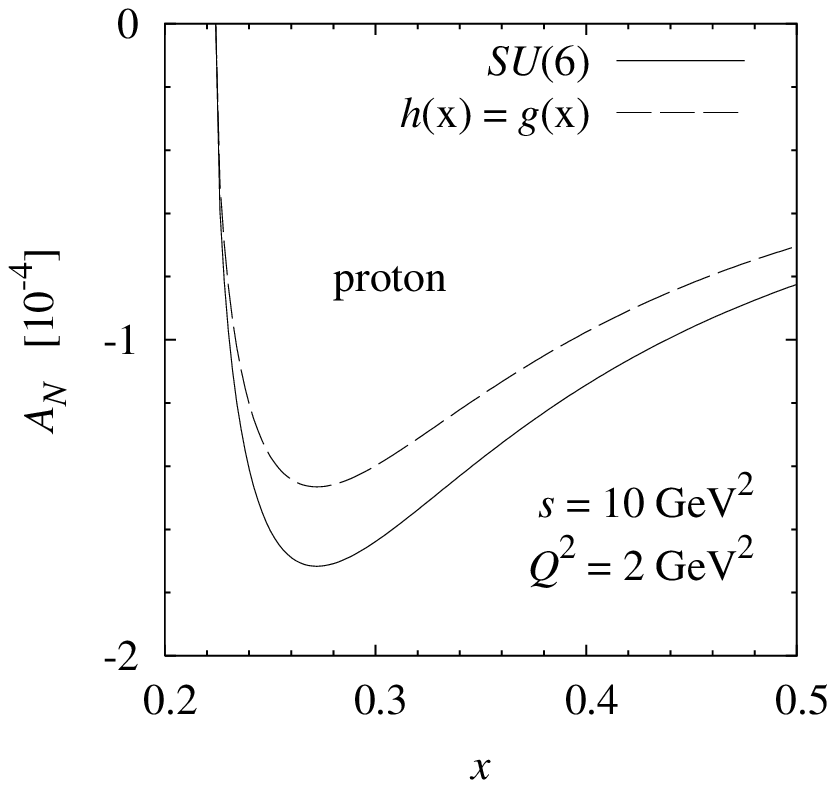}
& 
\includegraphics[width=0.49\textwidth]{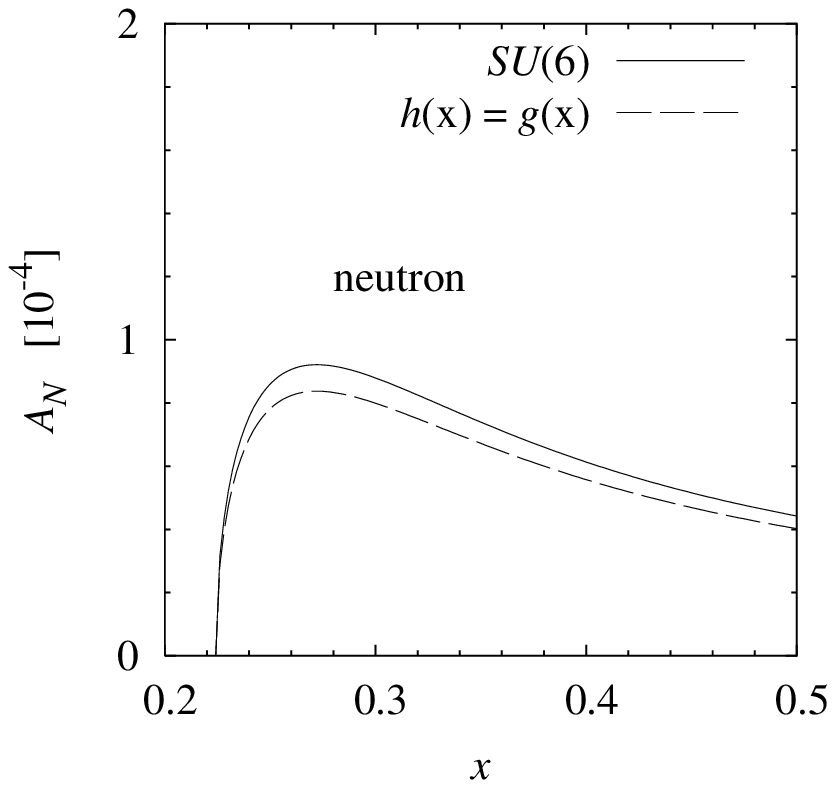}
\end{tabular}
\caption[]{The normal target spin asymmetry $A_N$ obtained
from the light--front constituent quark model with the
the composite nucleon approximation, Eq.~(\ref{neglect_k_T}), 
with different assumptions about the spin--flavor wave function:
(a)~$SU(6)$ symmetry, (b)~transversity = helicity distributions, 
$h(x) = g(x)$. The kinematics corresponds approximately to the planned 
Jefferson Lab Hall~A experiment ($s = 2 E_{\textrm{beam}} M + M^2$ 
is the squared electron--nucleon center--of--mass energy).}
\label{fig:comp_pn}
\end{figure}
Physically, this corresponds to a picture of the nucleon as an assembly 
of weakly interacting, massive quarks. In this 
picture the quark helicity--flip contribution to the transverse spin asymmetry 
(which is proportional to $M_q$) dominates over the quark 
helicity--conserving one (proportional to $\langle \bm{k}_T^2 \rangle$)
and can be calculated in a relativistic impulse approximation 
with on--shell quarks, which manifestly preserves electromagnetic gauge
invariance and is free of QED collinear divergences.
We emphasize that Eq.~(\ref{neglect_k_T}) is a theoretical idealization
made in order to enable an internally consistent calculation of 
the two--photon exchange cross section, not necessarily a reflection of
actual nucleon structure. Figure~\ref{fig:comp_pn} shows the results 
for the normal target spin asymmetry, Eq.~(\ref{A_N}), obtained in 
this picture with two different assumptions about the nucleon spin--flavor 
wave function: (a)~$SU(6)$ symmetry, (b)~quark transversity = helicity 
distributions, $h(x) = g(x)$. One sees that the asymmetry is of the order 
of $10^{-4}$, with different sign for proton and neutron. The dependence 
on the kinematic variables is investigated further in 
Ref.~\refcite{Afanasev:2007ii}.

The normal spin asymmetry in electron--neutron DIS is to
be measured in a dedicated Jefferson Lab Hall~A experiment \cite{PR-07-013}
with a polarized ${}^3\text{He}$ target ($E_{\text{beam}} = 6 \, \text{GeV},
x = 0.1 - 0.45, Q^2 = 1 - 3.5 \, \text{GeV}^2$), with a projected
absolute statistical error of $\delta A_N = 2 \textrm{--} 4\times 10^{-4}$
in each of the four $Q^2$ bins. This measurement will improve 
the sensitivity of the only previous measurement at
SLAC \cite{Rock:1970sj} by two orders of magnitude (in the SLAC
experiment the asymmetry was found to be compatible with zero
at the level of $\sim 3.5\%$). The statistical error of the planned JLab 
measurement is of the same order as the value of the asymmetry predicted
by the composite nucleon approximation Eq.~(\ref{neglect_k_T}), 
see Fig.~\ref{fig:comp_pn}. However, as already stated, this approximation 
was made for technical reasons and can provide only a rough estimate 
of the expected asymmetry. More realistic dynamical model calculations
of this observable are certainly needed. Work on this problem is 
in progress.
\bibliographystyle{ws-procs9x6}

\begin{thebibliography}{10}
%
%
\bibitem{Miller:2002qb}
G.~A. Miller and M.~R. Frank, {\em Phys. Rev.} {\bf C65}, 065205 (2002).
%
%
\bibitem{Miller:2004rc}
G.~A. Miller, {\em Phys. Rev.} {\bf C69}, 052201 (2004).
%
%
\bibitem{Afanasev:2007ii}
  A.~Afanasev, M.~Strikman and C.~Weiss,
  arXiv:0709.0901 [hep-ph].
%
%
\bibitem{Afanasev:2004hp}
A.~V.~Afanasev and N.~P.~Merenkov,
{\em Phys. Lett.} {\bf B599}, 48 (2004).
%
%
\bibitem{Metz:2006pe}
A.~Metz, M.~Schlegel and K.~Goeke, {\em Phys. Lett.} {\bf B643}, 319 (2006).
%
%
\bibitem{Anthony:2002hy}
P.~L. Anthony {\em et~al.}, {\em Phys. Lett.} {\bf B553}, 18 (2003).
%
%
\bibitem{Zheng:2004ce}
X.~Zheng {\em et~al.}, {\em Phys. Rev.} {\bf C70}, 065207 (2004).
%
%
\bibitem{Radyushkin:2000jy}
A.~V. Radyushkin and C.~Weiss, {\em Phys. Lett.} {\bf B493}, 332 (2000).
%
%
\bibitem{Radyushkin:2000ap}
A.~V. Radyushkin and C.~Weiss, {\em Phys. Rev.} {\bf D63}, 114012 (2001).
%
%
\bibitem{Penttinen:2000dg}
M.~Penttinen, M.~V. Polyakov, A.~G. Shuvaev and M.~Strikman, {\em Phys. Lett.}
  {\bf B491}, 96 (2000).
%
%
\bibitem{PR-07-013}
X.~Jiang {\em et~al.}, Target normal single--spin asymmetry in inclusive {DIS}
  $n(e,e')$ with a polarized ${}^3\text{He}$ target, 
  Jefferson Lab Hall~A Experiment E--07--013.
%
%
\bibitem{Rock:1970sj}
S.~Rock {\em et~al.}, {\em Phys. Rev. Lett.} {\bf 24}, 748 (1970).
%
%
\end{thebibliography}
\end{document}